\documentclass[pre,twocolumn,floatfix,showpacs,superscriptaddress]{revtex4}
\usepackage{amsmath}
\usepackage{graphicx}
\usepackage{psfrag}
\usepackage[outercaption]{sidecap}

\begin{document}
\title{Parallel-tempering cluster algorithm for computer simulations of critical phenomena}
\author{Elmar~Bittner}
\email{E.Bittner@thphys.uni-heidelberg.de}\affiliation{Institut f\"ur Theoretische Physik and Centre for
Theoretical Sciences (NTZ) -- Universit\"at Leipzig,\\
Postfach 100\,920, D-04009 Leipzig, Germany}
\affiliation{Institut f\"ur Theoretische Physik -- Universit\"at Heidelberg, Philosophenweg 16,
D-69120 Heidelberg, Germany}
\author{Wolfhard~Janke}
\email{Wolfhard.Janke@itp.uni-leipzig.de}
\affiliation{Institut f\"ur Theoretische Physik and Centre for
Theoretical Sciences (NTZ) -- Universit\"at Leipzig,\\
Postfach 100\,920, D-04009 Leipzig, Germany}
\date{\today}

\begin{abstract}
In finite-size scaling analyses of Monte Carlo 
simulations of second-order phase transitions one often needs
an extended temperature range around the critical point. By
combining the parallel tempering algorithm with cluster updates and
an adaptive routine to find the 
temperature window of interest, we introduce a flexible and powerful method
for systematic investigations of critical phenomena.
As a result, we gain one to two orders of magnitude in the performance for 2D and 3D Ising models
in comparison with the recently proposed Wang-Landau recursion for cluster algorithms
based on the multibondic algorithm,
which is already a great improvement over the standard multicanonical variant.

\end{abstract}
\maketitle

%%%%%%%%%%%%%%%%%%%%%%%%%%%%%%%%%%%%%%%%%%%%%%%%%%%%%%%%%%%%%%%%%%%%%%

\section{Introduction}

While much attention has been paid in the past to simulations of
first-order phase transitions and systems with rugged free-energy landscapes
in generalized ensembles (umbrella, multicanonical, Wang-Landau, 
parallel/simulated tempering sampling) \cite{rugged}, the merits of this 
non-Boltzmann sampling approach also for simulation studies of critical
phenomena have been pointed out only recently. In Ref.~\cite{bbwj}, Berg and 
one of the authors combined multibondic sampling \cite{JaKa95} with the
Wang-Landau recursion \cite{WaLa01} to cover the complete ``desired'' critical
temperature window in a single simulation for each lattice size, where the 
``desired'' range derives from a careful finite-size scaling (FSS) analysis
of all relevant observables. 
% Since the individual reweighting ranges of the
% involved observables may be quite disparate, a careful scaling analysis is an
% % the second 
% important ingredient of the method. 

Recent developments in the field of Graphic Processing Units (GPUs) make it possible to 
have access to a massively parallel computing solution at a low cost. To use these devices in a
most efficient way new parallelized algorithms are needed. 
Our parallel tempering cluster algorithm is a combination of 
replica-exchange methods~\cite{pt} with the Swendsen-Wang cluster 
algorithm~\cite{sw} and
is therefore predestinated for use in such devices.

\section{Parallel-Tempering Cluster Algorithm}

For the parallel-tempering procedure 
of the combined algorithm
we use a set 
of $N_{\rm rep}$ replica, where the number of replica depends on the 
% reweighting range %??? that is
``desired'' range that is
needed for the FSS analysis~\cite{opti}. 
To determine this range we perform at the beginning of our simulations
a short run in a reasonable temperature interval.
We choose the number of replica $N_{\rm rep}$ such that the acceptance rate $A(1 \rightarrow 2)$
between adjacent replica is about 50\%, which can be calculated from  
\begin{equation}
        \label{acc}
        A(1 \rightarrow 2)=\!\!\sum_{E_1,E_2} P_{\beta_1}(E_1) P_{\beta_2}(E_2) 
         P_{\rm PT}(E_1,\beta_1 \rightarrow E_2,\beta_2),
 \end{equation}
where $P_{\beta_i}(E_i)$ is the probability for replica $i$ at inverse temperature $\beta_i$ to have energy $E_i$ and
\begin{equation}
        \label{prob}
        P_{\rm PT}(E_1,\beta_1 \rightarrow E_2,\beta_2)=\min [1,\exp(\Delta \beta \Delta E)]
 \end{equation}
%
% $ P_{\rm PT}(E_1,\beta_1 \rightarrow E_2,\beta_2)$ 
with $\Delta \beta=\beta_2-\beta_1$, $\Delta E=E_2-E_1$ 
is the probability to accept a proposed exchange 
of different, usually adjacent replica.  
% This probability can be written as 
% where $\Delta \beta=\beta_2-\beta_1$ is the difference between the inverse temperatures of the two replica and
% $\Delta E=E_2-E_1$ their energy difference.
This choice of $A(1 \rightarrow 2)$ ensures that the multi-histogram reweighting~\cite{multi} 
works properly and the flow in (inverse) temperature space, that is, 
the rate of round trips between low and high temperatures is optimal~\cite{opti}.
This is the main difference to our preliminary note~\cite{pos}, where we used histogram overlaps instead.
Using the data of this short run as input for the 
multi-histogram reweighting routine, we determine the pseudo-critical points $C^{\rm max}=C(\beta^{\rm max}_C)$ of 
the specific heat $C(\beta)=\beta^2 V (\langle e^2\rangle - \langle e\rangle^2)$ and $\chi^{\rm max}$ of
the susceptibility $\chi(\beta)=\beta V (\langle m^2\rangle - \langle |m|\rangle^2)$, where
$e=E/V$ is the energy density, $m=M/V$ the magnetization density, 
% with $V = L^d$ denoting the size of the system. Furthermore, we 
and $V = L^d$ the size of the system. Furthermore, we 
measured the maxima of the slopes of the magnetic cumulants, $U_{2k}(\beta)=1-\langle m^{2k}\rangle/3\langle |m|^k\rangle^2$ ($k=1,2$),
% and $U_4(\beta)=1-\langle m^4\rangle/3\langle m^2\rangle^2$, 
and of the derivatives of the
magnetization, $d\langle |m|\rangle/d\beta$, $d\langle \ln |m|^k\rangle/d\beta$ ($k=1,2$),
%, and $d\langle \ln m^2\rangle/d\beta$, 
respectively. 
We also include the first structure factor $S_{k_1}$ (see, e.g., Ref.~\cite{stanly}) 
in our measurement scheme to allow a direct comparison with the results of Ref.~\cite{bbwj}.

\begin{figure}[t]
\centering
\includegraphics[scale=0.878]{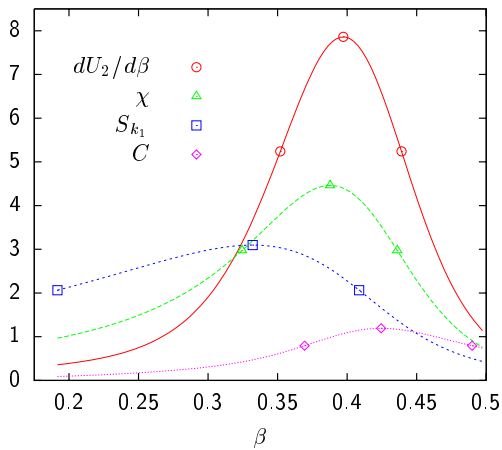}
\caption{\label{betafig}(Color online) Reweighted observables for the 2D Ising model with $L=8$. The symbols mark the maximum values 
$S^{\rm max}$ and the value  $S(\beta^{+/-}_S)=r S^{\rm max}$ with $r=2/3$.}
\end{figure}

Next we determine $\beta$ values where the observables $S=\{C,\chi,\dots \}$ reach the
value $S(\beta^{+/-}_S)=r S^{\rm max}$ with $r \le 1$.
% , where we use the moderate value $r=2/3$.  
This leads to a sequence of $\beta^{+/-}_S$ values satisfying 
$\beta^-_S \le \beta^{\rm max}_S$, 
and 
$\beta^+_S \ge \beta^{\rm max}_S$ 
as illustrated for the two-dimensional (2D) Ising model in Fig.~\ref{betafig}. 
% In Fig.~\ref{betafig}, we show as an example for such a sequence
% the reweighted curves for $C$, $\chi$, $dU_2/d\beta$, %$d\langle \ln |m|\rangle/d\beta$ 
% and $S_{k_1}$ for the two-dimensional (2D) Ising model with linear lattice size $L=8$. 
The actual simulation range is then given by the largest interval 
% of the sequence of all $\beta_S^{+/-}$ values. 
covered by these $\beta_S^{+/-}$ values, i.e., for the example in Fig.~\ref{betafig}
the ``desired'' simulation window would be $[\beta_{S_{k_1}}^-,\beta^+_C]$.
% In our example in Fig.~\ref{betafig}, this would lead to the ``desired'' range $[\beta_{S_{k_1}}^-,\beta^+_C]$.
% As one also can see in this figure, the value of $\beta_{S_{k_1}}^-$ is further away from the critical point 
% than all other $\beta_S^{+/-}$ values; therefore, if one is not particularly interested in the first structure factor,
% then the ``desired'' simulation range can be chosen narrower. 
%We now use the thus determined interval with the same number of replica for our actual measurement run. This interval is usually 
%narrower then the original estimate, hence the overlap of the histograms becomes larger and the 
%applicability of the multi-histogram reweighting method is assured. 

In practice we start by estimating for a very small system a reasonable 
(inverse) temperature interval $[\beta^-,\beta^+]$ and the number of replica
$N_{\rm rep}$ by trial and error.
%
%Let us now illustrate the work flow of our new method for the 2D Ising model.
%Here we started with a reasonable choice of the (inverse) temperature interval,
%$\beta^-=0.15$ and $\beta^+=0.6$, and the number of replica $N_{\rm rep} = 7$ for our smallest lattice $L=8$. 
For successively larger systems 
we use the measured temperature interval and $N_{\rm rep}$ of the next smaller system as input parameters. The work flow of our method is then
given by the following
%Then we used the following
general recipe:
\begin{enumerate}
%\item use the temperature interval and the number of replica of the next smaller system,
\item\label{step1} compute the simulation temperatures of the replica equidistant in $\beta$, 
\item perform several hundred thermalization sweeps and a short measurement run,
%\item perform a short measurement run,
\item check the histogram overlap between adjacent replica: if the overlap is 
too small ($< 10\%$), add a further replica and go back to step~\ref{step1}, else go on, 
\item use multi-histogram reweighting to determine $\beta_S^-$ and $\beta_S^+$ for all observables $S$,
leading to the temperature interval $[\beta_{\rm min}^-,\beta^+_{\rm max}]=[\min_S \{\beta_{S}^-\}, \max_S \{\beta_{S}^+\}]$,
\item\label{step2} start with $\beta^-=\beta_{\rm min}^-$ and compute a sequence of temperatures $\beta_i$ with fixed acceptance rate $A(1 \rightarrow 2)$ until
$\beta_i=\beta^+ \ge \beta^+_{\rm max}$,
\item perform several thousand thermalization sweeps and a long measurement run.
%\item perform the measurement run.
\end{enumerate}

\begin{table}[b]
\caption{Simulation windows and numbers of replica (see text) for the 2D Ising model simulations with $r=2/3$ on $L^2$ lattices.\label{tab2d}}
\centering
\begin{tabular}{rccccc}
\hline \hline
\makebox[0.7cm][r]{$L$}& \makebox[2cm][c]{$\beta^{-}=\beta_{S_{k_1}}^-$} & \makebox[2cm][c]{$\beta_{\rm max}^+=\beta_{C}^+$}& \makebox[2cm][c]{$\beta^{+}$} &
\makebox[0.7cm][c]{$N_{\rm rep}$} & \makebox[0.7cm][c]{$N_{\rm rep}^*$ }\\ \hline
8    & 0.191\,636 & 0.489\,877  & 0.565\,730 & 7 & 6  \\
16   & 0.320\,580 & 0.470\,826  & 0.471\,618 & 7 & 7  \\
32   & 0.380\,475 & 0.460\,499  & 0.471\,780 & 9 & 7  \\
64   & 0.410\,832 & 0.452\,619  & 0.457\,003 & 10& 8  \\
128  & 0.425\,789 & 0.448\,345  & 0.449\,188 & 11& 8  \\
256  & 0.433\,267 & 0.446\,069  & 0.446\,392 & 13& 9  \\
512  & 0.437\,042 & 0.443\,171  & 0.443\,674 & 14& 10 \\
1024 & 0.438\,854 & 0.442\,400  & 0.442\,464 & 16& 10 \\ \hline 

$\infty$ &\multicolumn{3}{c}{$\beta_c=\ln(1+\sqrt{2})/2=0.440\,686\,7935\dots$} &&\\ \hline \hline\\

\end{tabular}
\end{table}

\section{Results}

\subsection{2D Ising model}

\begin{figure}[t]
\centering
\includegraphics[scale=0.878]{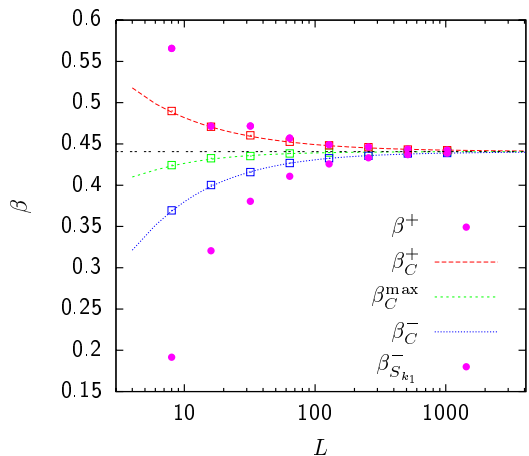}
\caption{\label{fig_c_beta}(Color online) The ``desired'' temperature interval for $r=2/3$ 
% determined using the specific heat i
as a function of the system size.
The horizontal line indicates the critical inverse temperature $\beta_c$, the other lines show 
exact results derived from Ref.~\cite{FF}.
% calculated from the
% formula of Ferdinand and Fisher~\cite{FF}. 
The circles indicate the
simulation windows determined fully automatically, cf.\ Table~\ref{tab2d}, and the boxes show
for completeness the measured values for $\beta_C^+$, $\beta_C^{\rm max}$ and $\beta_C^-$.}
\end{figure}

%After choosing an input temperature interval and a number of replica for the smallest system, our program
Applying this recipe to the 2D Ising model, our computer program
simulated 
system sizes from  $L=8$ up to $L=1024$ fully automatically. This shows how robust our new method is.
Table~\ref{tab2d} gives an overview of the resulting
temperature intervals and the numbers of replica needed
when for comparison with Ref.~\cite{bbwj} $r=2/3$ is used. % which roughly scale with $L^{-1/\nu}$, where $\nu$ is the standard critical exponent of the correlation length.
Due to the acceptance rate criterion in step 5 of our iterative procedure, 
the upper bounds $\beta^{+}$ of the temperature intervals
slightly overshoot $\beta_{\rm max}^{+}$ and show relatively large fluctuations. 
With increasing system sizes this discretization effect becomes less pronounced, see
Fig.~\ref{fig_c_beta}
%% .
%% %This scaling can also be used to extrapolate the input interval for larger system sizes.
%% In two dimensions, the branch coming from the specific heat 
%% exhibits a multiplicative logarithmic term, therefore, one could use this knowledge
%% to improve the extrapolation for this special case. We refrain from
%% such modifications to keep the program as generally usable as
%% possible.
where we compare 
%% For comparison we show in Fig.~\ref{fig_c_beta} 
the automatically determined interval of our algorithm
with
the exact
temperature interval $[\beta_C^-,\beta^+_C]$ using the
specific-heat formula of Ferdinand and Fisher~\cite{FF}.
%% and
%% In Table~\ref{tab2d} we also show the numbers of replica for the given input interval. The fact that 
%% we start with $\beta^-=\beta_{S_{k_1}}^-$ to calculate the simulation temperatures using a constant acceptance rate between adjacent 
%% temperatures and the condition that the largest $\beta_i=\beta^+$ of the sequence has to be at least equal to $\beta^+_{\rm max}=\beta_C^+$ determines this number. 

\begin{figure}[t]
\centering
\includegraphics[scale=0.878]{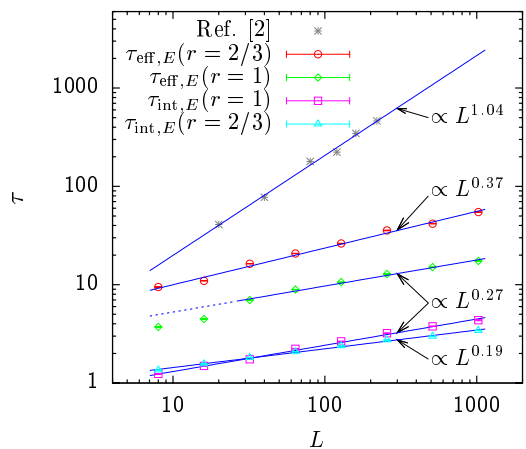}
\caption{\label{tau2d}(Color online) Autocorrelation times $\tau_{\rm int}$ and $\tau_{\rm eff}$ for the 
energy of the 2D Ising model, where $\tau_{\rm eff}=N_{\rm rep} \, \tau_{\rm int}$ 
with $N_{\rm rep}$ denoting the number of replica, cf.\ Table~\ref{tab2d}.}
\end{figure}

To assess the performance of the method, we measured the integrated autocorrelation
times $\tau_{\rm int}(\beta_i,L)$ and determined the maximum over all replica,  
$\tau_{\rm int}(L)=\max_{i=1,N_{\rm rep}}\tau_{\rm int}(\beta_i,L)$, for each 
lattice size. By fitting the critical slowing down ansatz $\tau_{\rm int}(L) \propto L^z$ 
to the data, we find for 
% of the maximal integrated autocorrelation times $\tau_{\rm int}(L)=\max_{i=1,N_{\rm rep}}\tau_{\rm int}(\beta_i,L)$ 
% for the energy, squared magnetization and the first structure factor
$e$, $m^2$, and $S_{k_1}$ rather small dynamical critical exponents
$z=0.19(1)$, $z=0.11(1)$, and $z=0.01(1)$, respectively.
%
%scale only weakly with the system size $L$. 
As an example, we compare in Fig.~\ref{tau2d} our data for $\tau_{{\rm int},E}(L)$ 
of the energy with the results of Ref.~\cite{bbwj}.
% The critical slowing down scales $\propto L^z$ with dynamical critical exponent $z=0.19(1)$. 
% To make a fair comparison with other methods, we also take the computational effort into account
Of course, in our case the 
computational effort depends linearly on the number of replica needed.
We therefore also show 
% in Fig.~\ref{tau2d} also 
the 
% into account
% and include the number of replica $N_{\rm rep}$ into the definition of an 
effective autocorrelation time 
% according to 
$\tau_{\rm eff}=N_{\rm rep} \, \tau_{\rm int}$, which enables a fair
comparison in units of lattice sweeps. We see that for $L > 100$, our 
$\tau_{\rm eff}$ is more than one order of magnitude smaller than using 
the recently proposed multibondic Wang-Landau method~\cite{bbwj}.
% than using the recently proposed multibondic Wang-Landau method~\cite{bbwj}.
For $r=2/3$, $N_{\rm rep}$ grows 
with increasing lattice size
$\propto L^{z'}$ with $z' \approx 0.18$ (cf.\ Table~\ref{tab2d}).
Consequently, $\tau_{\rm eff} \propto L^{z_{\rm eff}}$ with
$z_{\rm eff} = z + z'$. For the energy this gives $z_{\rm eff} \approx 0.37$,
which is still much smaller than the exponent $z \approx 1.04$ found in 
Ref.~\cite{bbwj},
% with the recently proposed multibondic Wang-Landau method~\cite{bbwj}, 
so that the gain in efficiency becomes more and more pronounced with increasing 
lattice size.

%$\propto L^{z_{\rm eff}}$, yielding for the energy an exponent
%$z_{\rm eff}=0.37(1)$, i.e., $N_{\rm rep} \propto L^{0.18}$.
%Similarly, we find $z=0.11(1)$ and $z_{\rm eff}=0.29(1)$ for $m^2$ and $z=0.01(1)$ and $z_{\rm eff}=0.19(1)$ for $S_{k_1}$.
%For $\tau_{\rm int}$ and $\tau_{\rm eff}$ of $m^2$ we find slightly smaller values, $z=0.11(1)$ and $z_{\rm eff}=0.29(1)$, respectively. 
%For $S_{k_1}$ the dynamical critical exponent is $z=0.01(1)$ and for $\tau_{\rm eff}$ we find $z_{\rm eff}=0.19(1)$.
%Even the larger absolute values of the effective autocorrelation times are almost an order of magnitude smaller
%and scale with a much smaller exponent than  using the recently proposed multibondic Wang-Landau method~\cite{bbwj}.

The choice $r=2/3$ of Ref.~\cite{bbwj} is quite conservative as even for $r=1$ 
the peaks determining the left and right boundary of the ``desired'' simulation window
are usually sufficiently well sampled. This amounts to a somewhat smaller temperature
range and repeating the above procedure with $r=1$, we arrive at a smaller
$z_{\rm eff} = z = 0.27$, see Fig.~\ref{tau2d}. Here 
$z_{\rm eff} = z$ because the integer valued $N_{\rm rep}$ turn out be so 
small (3 for $L=8,16$ and 4 for $ 32 \le L \le 1024$) that they practically stay
constant over a wide range of system sizes.
% (since $N_{\rm rep}$ can only take integer values). 
For large $L$, this leads to a significant gain, and also for
the moderate system sizes of Fig.~\ref{tau2d}, $\tau_{\rm eff}$ is already 
reduced by a factor of about 3 compared with the case of $r=2/3$ and a factor
of about 100 compared with Ref.~\cite{bbwj}.

On the other hand, for $r < 2/3$, the ``desired'' simulation window (and hence also 
$N_{\rm rep}$) becomes larger, and the effective autocorrelation time 
$\tau_{\rm eff}$ grows faster with system size. For example for $r=1/2$, we 
find $z' \approx 0.32$. With $z=0.23(1)$ this leads to $z_{\rm eff}=0.55(2)$ 
for the energy and similar results for $m^2$ and $S_{k_1}$.

\begin{figure}[t]
\centering
\includegraphics[scale=0.88]{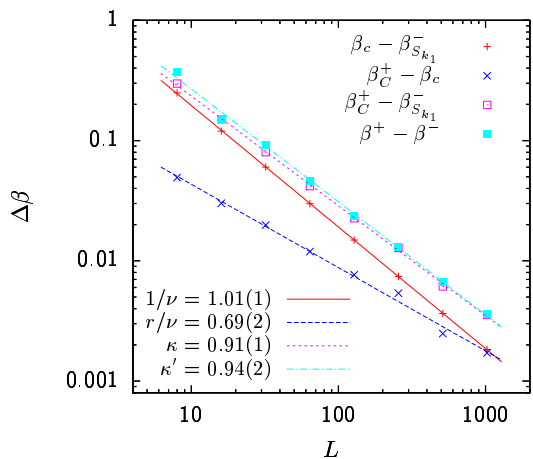}
\caption{\label{delta_beta} 
(Color online) FSS of the ``desired'' simulation window for the 2D Ising model with $r=2/3$.
%and $r=2/3$, where $\beta_c$ denotes the critical point.
}
\end{figure}

In order to understand the FSS of $\tau_{\rm eff}$ theoretically, we have to 
recall that the 2D Ising model is a particular case due to
its logarithmic divergence of the specific heat $C$
(with critical exponent $\alpha = 0$).
First, this leads to a logarithmic correction in the FSS of the reweighting 
range, $\Delta \beta_{\rm rew} \propto L^{-1/\nu}/\sqrt{\ln L}$. Second, 
when $C$ is included in the
``desired'' quantities, it (empirically) determines the upper bound 
$\beta_{\rm max}^{+}=\beta_C^{+}$, which 
then does not scale generically with $L^{-1/\nu}$ but with $L^{-r/\nu}$ \cite{bbwj1}.
For the ``desired'' range 
% $\Delta \beta_{\rm desired} = 
$\beta^+-\beta^- \approx \beta_C^{+} -\beta_{S_{k_1}}^{-} 
= a L^{-1/\nu} + b L^{-r/\nu}$, this shows that $L^{-r/\nu}$ is 
asymptotically the leading term for $r<1$, so that 
$N_{\rm rep} = (\beta^+-\beta^-)/\Delta \beta_{\rm rew} 
\rightarrow L^{(1-r)/\nu} \sqrt{\ln L}$ for $L \gtrsim L_\times =(a/b)^{\nu/(1-r)}$. 
However, as Fig.~\ref{delta_beta} demonstrates for $r=2/3$, this is not observed for
practically accessible lattice sizes: Instead of $L^{-2/3}$ (with $\nu = 1$) we 
observe 
% $\Delta \beta_{\rm desired} 
$\beta^+-\beta^- \propto L^{-\kappa'}$ with $\kappa' = 0.94$. 
The reason is the very slow crossover
from subleading to leading scaling behavior. In fact, whereas both 
$\beta_c - \beta_{S_{k_1}}^- \propto L^{-1.01}$ and 
$\beta_C^+ - \beta_c \propto L^{-0.69}$ do scale as expected, 
the asymptotically leading term starts to dominate only at around $L=1000 \approx L_\times$
with $a/b = 10$, which implies 
$N_{\rm rep} \propto L^{0.06} \sqrt{\ln L}$ for $L \lesssim 1000$. 
Since in this range to a very good approximation
$\sqrt{\ln L} \approx L^{0.11}$ (which can be easily verified directly),
this effectively leads to 
% $N_{\rm rep} \propto  L^{z'}$
% with $z'=0.06+0.11=0.17$, in good agreement with
$N_{\rm rep} \propto  L^{0.17}$,
in good agreement with
our direct estimate $z' \approx 0.18$.
For $r=1/2$, $L^{-1}$ and $L^{-r} = L^{-1/2}$
differ more strongly and we observe the crossover in the FSS of $\beta^+-\beta^-$
already at about $L = 150 \approx L_\times$ with $a/b=12$, yielding here
$\beta^+-\beta^- \propto L^{-0.80}$ and hence
% $N_{\rm rep} \propto L^{0.20} \sqrt{\ln L} \approx L^{z'}$
% with $z' = 0.20+0.11 = 0.31$, again as observed above.
$N_{\rm rep} \propto L^{0.20} \sqrt{\ln L} \approx L^{0.31}$,
again as fitted directly.
% which explains the difference between $z$ and the effective exponent $z_{\rm eff}$.
Only for $r=1$, $\beta^+-\beta^- \propto L^{-1/\nu}$ and 
$N_{\rm rep} \propto \sqrt{\ln L}$, which 
%varies only
%by a factor $1.8$ from $L=8$ to 1024.
as explained above is hardly visible for small (integer) values of $N_{\rm rep}$.

If one omits $C$ as a criterion to specify the ``desired'' FSS window,
both the upper bound $\beta^+$ and
$\beta^+-\beta^-$ scale $\propto L^{-1/\nu}$ for any value of $r$, so that 
$N_{\rm rep}^* \propto \sqrt{\ln L}$, see the last column in Table~\ref{tab2d} for the
case $r=2/3$.
If the simulation window is now too narrow to determine the critical exponent $\alpha$ directly,
one still can use the hyperscaling relation $\alpha=2-d\nu$ with $d$ the 
dimensionality.
% For these temperature intervals, the ``bare'' dynamical critical exponents $z$
Whereas the ``bare'' dynamical critical exponents $z$ do not differ much,
due to the smaller number of replica needed, 
$\tau_{\rm eff} = N_{\rm rep}^* \, \tau_{\rm int}$ is slightly 
smaller than in the case including $C$. 
If one again simply fits a
power law to both $\tau_{\rm int}$ and $\tau_{\rm eff}$ (i.e., 
ignores the logarithmic behavior of $N^*_{\rm rep}$), one finds excellent fits with 
$z=0.20(1)$ and $z_{\rm eff}=0.31(1) = z + 0.11$ for the energy 
($z=0.11(1)$, $z_{\rm eff}=0.22(1)$ for $m^2$ and
$z=0.03(1)$, $z_{\rm eff}=0.13(1)$ for $S_{k_1}$), 
confirming that effectively $\sqrt{\ln L} \approx L^{0.11}$ 
for moderately large $L \lesssim 1000$. 
% (which can be easily verified directly). 
% This also explains why we observe
% in the first case (with $\beta^+_C$ determining the upper bound $\beta^+$)
% effectively $N_{\rm rep} \propto L^{0.06} \sqrt{\ln L}
% \approx L^{0.06+0.11} = L^{0.17}$. 

% We also explored the dependence of $z$ and $z_{\rm eff}$
% % the dynamical critical exponent 
% on the parameter $r$.
% By varying $r$, the ``bare'' dynamical critical exponent $z$ 
% % remains the same. 
% stays almost constant. However, 
% the effective autocorrelation time $\tau_{\rm eff}$ grows faster for smaller values of 
% $r$, i.e., a broader reweighting range. 
% For example for $r=0.5$, we find 
% $z=0.23(1)$ and $z_{\rm eff}=0.55(2)$ for the energy and similar results for $m^2$ and $S_{k_1}$.  
% Here $L^{-1}$ and $L^{-r} = L^{-1/2}$ 
% differ more strongly and the crossover in the FSS of $\Delta \beta_{\rm desired}$ happens 
% already at about $L = 150 \approx L_\times$ with $a/b=12$, yielding effectively 
% $\Delta \beta_{\rm desired} \propto L^{-0.80}$.
% This implies $N_{\rm rep} \propto L^{0.20} \sqrt{\ln L} \approx L^{0.20+0.11} = L^{0.31}$,
% which explains the difference between $z$ and the effective exponent $z_{\rm eff}$.

\begin{table}[b]
\caption{Simulation windows and numbers of replica for the 3D Ising model 
simulations with $r=2/3$ on $L^3$ lattices.\label{tab3d}}
\centering
\begin{tabular}{rrrrr}
\hline \hline
\makebox[0.7cm][r]{$L$}& \makebox[1.8cm][c]{$\beta^{-}=\beta_{S_{k_1}}^-$} & \makebox[1.8cm][c]{$\beta_C^{+}$} &\makebox[1.8cm][c]{$\beta^{+}$} & \makebox[0.7cm][r]{$N_{\rm rep}$} \\ \hline
  4 &  0.061\,955 & 0.284\,066 & 0.311\,321 &  7\\ 
  6 &  0.142\,959 & 0.259\,876 & 0.272\,522 &  8\\ 
  8 &  0.173\,712 & 0.252\,866 & 0.259\,229 &  9\\ 
 10 &  0.188\,447 & 0.246\,916 & 0.252\,136 & 10\\ 
 12 &  0.197\,404 & 0.241\,159 & 0.243\,106 & 10\\ 
 16 &  0.206\,422 & 0.236\,610 & 0.236\,773 & 11\\ 
 20 &  0.211\,183 & 0.233\,407 & 0.233\,621 & 12\\ 
 24 &  0.213\,807 & 0.232\,462 & 0.233\,016 & 14\\ 
 28 &  0.215\,553 & 0.229\,853 & 0.230\,301 & 14\\ 
 32 &  0.216\,670 & 0.229\,221 & 0.229\,293 & 15\\ 
 36 &  0.217\,572 & 0.228\,026 & 0.228\,613 & 16\\ 
 40 &  0.218\,172 & 0.227\,703 & 0.228\,025 & 17\\ 
 48 &  0.219\,082 & 0.226\,669 & 0.226\,754 & 18\\ 
 56 &  0.219\,621 & 0.225\,353 & 0.225\,555 & 18\\ 
 64 &  0.220\,031 & 0.224\,758 & 0.224\,775 & 18\\ 
 72 &  0.220\,309 & 0.224\,359 & 0.224\,435 & 19\\ 
 80 &  0.220\,505 & 0.224\,331 & 0.224\,347 & 21\\

%  4 &  0.061\,955 & 0.242\,829 & 0.284\,066 &  7\\ 
%  6 &  0.142\,959 & 0.233\,731 & 0.259\,876 &  8\\ 
%  8 &  0.173\,712 & 0.229\,560 & 0.252\,866 &  9\\ 
% 10 &  0.188\,447 & 0.228\,032 & 0.246\,916 & 10\\ 
% 12 &  0.197\,404 & 0.226\,155 & 0.241\,159 & 10\\ 
% 16 &  0.206\,422 & 0.224\,854 & 0.236\,610 & 11\\ 
% 20 &  0.211\,183 & 0.223\,887 & 0.233\,407 & 12\\ 
% 24 &  0.213\,807 & 0.223\,165 & 0.232\,462 & 14\\ 
% 28 &  0.215\,553 & 0.223\,011 & 0.229\,853 & 14\\ 
% 32 &  0.216\,752 & 0.222\,686 & 0.228\,981 & 16\\ 
% 36 &  0.217\,572 & 0.222\,499 & 0.228\,026 & 16\\ 
% 40 &  0.218\,172 & 0.222\,396 & 0.227\,703 & 17\\ 
% 48 &  0.219\,082 & 0.222\,204 & 0.226\,669 & 18\\ 
% 56 &  0.219\,621 & 0.222\,037 & 0.225\,353 & 18\\ 
% 64 &  0.220\,031 & 0.221\,981 & 0.224\,758 & 18\\
% 72 &  0.220\,308 & 0.224\,358 & 0.224\,434 & 19\\
% 80 &  0.220\,505 & 0.224\,330 & 0.224\,347 & 21\\
\hline  %\hline
$\infty$ & \multicolumn{3}{c}{$\beta_c \simeq 0.221\,654\,59$~(Ref.~\cite{bloete})} &  \\ \hline \hline
\end{tabular}
\end{table}

\subsection{3D Ising model}

\begin{figure}[t]
\centering
\includegraphics[scale=0.878]{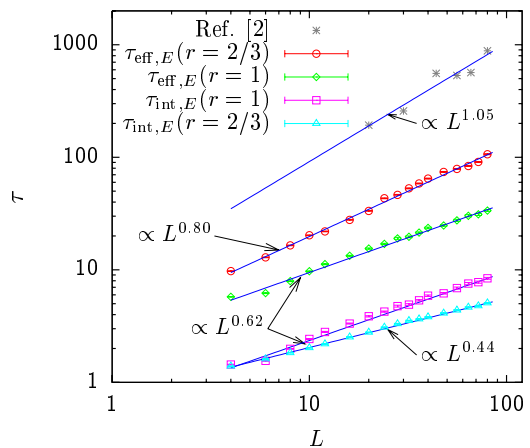}
\caption{\label{tau3d}(Color online) Same as Fig.~\ref{tau2d} for the 3D Ising model, cf.\ Table~\ref{tab3d}.
%
%Autocorrelation times \label{tau3d}$\tau_{\rm int}$ and $\tau_{\rm eff}$ for the energy of the 3D Ising model, where 
%$\tau_{\rm eff}=N_{\rm rep}\times \tau_{\rm int}$ and $N_{\rm rep}$ is the number of replica, cf. Table~\ref{tab3d}.}
}
\end{figure}

In the 3D Ising model where $\alpha  \approx 0.11 > 0$, both the 
reweighting range and the ``desired'' temperature window should scale
with $L^{-1/\nu}$, so that one would
expect that our routine will use the same number of replica for all system sizes. 
This is indeed the case for the simplest choice $r=1$, where our routine
determines for all lattice sizes $N_{\rm rep} = 4$. The autocorrelation analysis 
along the same lines as in 2D gives $z=z_{\rm eff}=0.62(2)$ for the energy
($z=z_{\rm eff} = 0.59(2)$ for $m^2$ and $z=z_{\rm eff} = 0.37(2)$ for $S_{k_1}$).
This exponent is thus again much smaller than $z \approx 1.05$ obtained in 
Ref.~\cite{bbwj}, and already for moderate system sizes $L \approx 40-80$
the values of $\tau_{\rm eff}$ are about $20-30$ times smaller, 
cf.\ Fig.~\ref{tau3d}.

If we choose $r=2/3$ as in Ref.~\cite{bbwj}, however,
% But as in the 2D case without the specific heat 
we find also here a weak system-size dependence 
% $N_{\rm rep}\propto L^{0.36(1)}$,
$N_{\rm rep}\propto L^{0.36}$,
cf.\ Table~\ref{tab3d} where also the automatically determined temperature intervals
are given. Here we obtain $z=0.44(1)$ and thus $z_{\rm eff}=0.80(1)$ for the 
energy, cf.\ Fig.~\ref{tau3d} 
($z=0.41(1)$, $z_{\rm eff}=0.78(1)$ for $m^2$ and $z=0.18(2)$, 
$z_{\rm eff}=0.55(2)$ for $S_{k_1}$).
The reason for this unexpected result can be 
traced back
to the fact
% This dependence can be explained by the fact 
that
% even by using the moderate value of $r=2/3$, 
the upper boundary $\beta^+_C$ of the ``desired'' simulation window, determined by
the low-temperature tail of $C$, lies for $r=2/3$ clearly outside the FSS region.
In fact, omitting $C$ as a criterion for the FSS window, $N_{\rm rep} = 7 - 9$ stays
almost constant.
The choice $r<1$ is thus less favorable, but even for $r=2/3$ one gains about
one order of magnitude in computing time compared with Ref.~\cite{bbwj}.

% To  cure this problem, we suggest to use the maximum position of observables directly, i.e.
% $r=1$, and therefore a slightly smaller interval $[\beta_{S_{k_1}}^{\rm max},\beta_C^{\rm max}]$ (or to omit $C$ as a criterion).
% In Table~\ref{tab3d} we also give an overview of the automatically determined temperature intervals 
% for the 3D Ising model 
% which are similar to the intervals compiled in Table~I of Ref.~\cite{bbwj}.
% A larger number of sweeps in the first short measurement run would lead to a better estimate for the
% temperature interval. 
% We used only about $1\%$ of our CPU time for this determination; increasing this
% percentage may gain a further improvement of the final results.
% In Fig.~\ref{tau3d} we show the integrated and effective autocorrelation times for the energy.
% Here we obtain for 
% the dynamical critical exponent $z=0.44(1)$ and $z_{\rm eff}=0.80(1)$. 
% We find $z=0.41(1)$ and $z_{\rm eff}=0.78(1)$ for $m^2$ and $z=0.18(2)$  and $z_{\rm eff}=0.55(2)$ for $S_{k_1}$.   
% In the 3D Ising model the absolute values of $\tau_{\rm int}$
% the integrated autocorrelation times 
% are almost two orders of magnitude smaller and even the $\tau_{\rm eff}$
% effective autocorrelation times
% are an order of magnitude smaller than those reported for the multibondic scheme in Ref.~\cite{bbwj}. 
% Since also the dynamical critical exponents are smaller, the asymptotic critical slowing down is less pronounced.
% 

\section{Conclusions}

To summarize, we have introduced a very flexible and simple approach for a systematic determination and simulation of
the critical temperature window of interest for second-order phase transitions, 
which one needs for an accurate estimation of critical exponents and other quantities characterizing
critical phenomena. 
The efficiency of the method depends of course on the chosen or available update scheme in the particular case, with non-local cluster flips being the
favorable choice.  
Since the setup of our method is completely general and can be combined also with 
any other update scheme 
(multigrid, worms, Metropolis, heat-bath, Glauber, $\dots$),
it could be employed for all simulations in 
statistical physics, chemistry and biology, 
high-energy physics and quantum field theory
where one is interested in critical phenomena.

%%%%%%%%%%%%%%%%%%%%%%%%%%%%%%%%%%%%%%%%%%%%%%%%%%%%%%%%%%%%%%%%%%%%%%%%%%

\section*{ACKNOWLEDGMENTS}
%\Acknowledgements 

Work supported by the Deutsche Forschungsgemeinschaft (DFG)
under grant Nos.~JA483/22-1/2 and 23-1/2, 
the EU RTN Network `ENRAGE' 
 -- {\em Random Geometry and Random Matrices: From Quantum Gravity to Econophysics\/}
under grant
No.~MRTN-CT-2004-005616, and by the computer-time grant No.~hlz10 of 
% the John von Neumann Institute for Computing (NIC),
NIC,
Forschungszentrum J\"ulich. 
% FZ J\"ulich. 
%%%%%%%%%%%%%%%%%%%%%%%%%%%%%%%%%%%%%%%%%%%%%%%%%%%%%%%%%%%%%%%%%%%%%%%%%%
%\vspace*{-1mm}

\end{document}